\documentclass[aps,prd,twocolumn, superscriptaddress]{revtex4}

\usepackage{amsmath}

\def\kappatr{\tilde{\kappa}_{tr}}

\begin{document}


\title{Erratum: New methods of testing Lorentz violation in electrodynamics}


\author{Michael Hohensee}
\thanks{Thanks David Phillips and Itay Yavin for helpful discussions.}
\author{Alex Glenday}
\affiliation{Department of Physics, Harvard University, Cambridge, Massachusetts 20138, USA}
\affiliation{Harvard-Smithsonian Center for Astrophysics, Cambridge, Massachusetts 02138, USA}
\author{Chih-Hao Li}
\affiliation{Harvard-Smithsonian Center for Astrophysics, Cambridge, Massachusetts 02138, USA}

\author{Michael Edmund Tobar}
\affiliation{School of Physics, the University of Western Australia, 35 Stirling Hwy, Crawley 6009, Western Australia, Australia}
\author{Peter Wolf}
\affiliation{LNE-SYRTE, CNRS, UPMC, Observatoire de Paris, 75014 Paris, France}


\date{\today}

\begin{abstract}
In an earlier paper \cite{Tobar:2005} (hep-ph/0408006), the bound on the Standard Model Extension photon-sector parameter $\tilde{\kappa}_{tr}$ \cite{Kostelecky:2002} (hep-ph/0205211) set by the heavy-ion storage-ring experiment of \citeauthor{Saathoff:2003} \cite{Saathoff:2003} was incorrectly reported.  We show that the correct bound on $\tilde{\kappa}_{tr}$ which resulted from this experiment is in fact $2.2\times 10^{-7}$.
\end{abstract}


\maketitle
In the original derivation of the leading order Lorentz-violating effects of $\kappatr$ to the Ives-Stilwell observable $\nu_{a}\nu_{p}/\nu_{0}^{2}$, second order and higher terms in $\beta_{at}=v_{at}/c$ were neglected \cite{Tobar:2005}.  In this limit, and apart from time-varying, frame-dependent factors of order unity, a nonzero $\kappatr$ produces a shift of the Ives-Stilwell observable that is proportional to $4\beta_{at}\beta_{lab}\kappatr$.   For terrestrial experiments, we take $\beta_{lab}\sim\beta_{\oplus}$, noting that the laboratory's velocity relative to the preferred (Sun-centered) frame is dominated by the Earth's orbital velocity, $\beta_{\oplus}\sim10^{-4}$ \footnote{There are also contributions to $\beta_{lab}$ from $\beta_{r}$, the component of the lab's motion due to the Earth's spin, but this is far smaller than $\beta_{\oplus}$.}.

For Saathoff and collaborators in \cite{Saathoff:2003}, however, $\beta_{at}\simeq 6\times10^{-2}$ is far larger than $\beta_{\oplus}$.  Because of this, we must consider terms proportional to $\beta_{at}^{2}$.  Noting that (equation (10) of \cite{Tobar:2005})
\begin{align}
\nu_{p}&= \nu_{0}\frac{\sqrt{1-\beta_{at}^{2}}}{1-\beta_{at}(c/c_{p})}, &\text{and } 
\nu_{a}&= \nu_{0}\frac{\sqrt{1-\beta_{at}^{2}}}{1+\beta_{at}(c/c_{a})},
\end{align}
then to second order in $\beta_{at}$,
\begin{equation}
\frac{\nu_{p}\nu_{a}}{\nu_{0}^{2}}=1+\beta_{at}\left(\frac{c}{c_{p}}-\frac{c}{c_{a}}\right) +\beta_{at}^{2}\left(\frac{c^{2}}{c_{a}^{2}}+\frac{c^{2}}{c_{p}^{2}}-\frac{c^{2}}{c_{a}c_{p}}-1\right).
\end{equation}
Further expanding the second order term to first order in $\kappatr$, we find that
\begin{equation}
\beta_{at}^{2}\left(\frac{c^{2}}{c_{a}^{2}}+\frac{c^{2}}{c_{p}^{2}}-\frac{c^{2}}{c_{a}c_{p}}-1\right)=2\beta_{at}^{2}\kappa_{tr}\label{eq:secondorderterm}
\end{equation}
with no time-variation due to the Earth's changing velocity in the Sun-centered frame.  Combining (\ref{eq:secondorderterm}) with the result of \cite{Tobar:2005} we obtain, to lowest order in $\kappa_{tr}$,
\begin{equation}
\frac{\nu_{p}\nu_{a}}{\nu_{0}^{2}}=1+2\kappa_{tr}\left[\beta_{at}^{2}+2\beta_{at}\beta_{\oplus}\left(\dots\right)\right] +O\left(\kappatr^{2}\right),
\end{equation}
where as specified in equation (12) of \cite{Tobar:2005} $\left(\dots\right)$ represents sinusoidally varying terms of order unity.  In the limit $\beta_{at}\gg\beta_{\oplus}$, the $\beta_{at}^{2}$ ``DC'' term dominates the time-dependent sidereal term.  Thus the previously derived equation (12) in \cite{Tobar:2005} is only valid when $\beta_{at}<\beta_{\oplus}$.  Upon comparison to the RMS \cite{Robertson:1949,Mansouri:1977} parameter $\hat{\alpha}$ (see equation (2) of \cite{Saathoff:2003}) bounded by \citeauthor{Saathoff:2003}, we find that in the limit that $\beta_{at}\gg\beta_{\oplus}$, a bound on $\hat{\alpha}$ is equivalent to a bound on $\kappatr$.  We therefore conclude that the limit set on $\kappatr$ by \cite{Saathoff:2003} is $\kappatr<2.2\times 10^{-7}$.


\bibliography{ErratumPRD71}

\begin{thebibliography}{5}
\expandafter\ifx\csname natexlab\endcsname\relax\def\natexlab#1{#1}\fi
\expandafter\ifx\csname bibnamefont\endcsname\relax
  \def\bibnamefont#1{#1}\fi
\expandafter\ifx\csname bibfnamefont\endcsname\relax
  \def\bibfnamefont#1{#1}\fi
\expandafter\ifx\csname citenamefont\endcsname\relax
  \def\citenamefont#1{#1}\fi
\expandafter\ifx\csname url\endcsname\relax
  \def\url#1{\texttt{#1}}\fi
\expandafter\ifx\csname urlprefix\endcsname\relax\def\urlprefix{URL }\fi
\providecommand{\bibinfo}[2]{#2}
\providecommand{\eprint}[2][]{\url{#2}}

\bibitem[{\citenamefont{Tobar et~al.}(2005)\citenamefont{Tobar, Wolf, Fowler,
  and Hartnett}}]{Tobar:2005}
\bibinfo{author}{\bibfnamefont{M.~E.} \bibnamefont{Tobar}},
  \bibinfo{author}{\bibfnamefont{P.}~\bibnamefont{Wolf}},
  \bibinfo{author}{\bibfnamefont{A.}~\bibnamefont{Fowler}}, \bibnamefont{and}
  \bibinfo{author}{\bibfnamefont{J.~G.} \bibnamefont{Hartnett}},
  \bibinfo{journal}{Phys. Rev. D} \textbf{\bibinfo{volume}{71}},
  \bibinfo{pages}{025004} (\bibinfo{year}{2005}).

\bibitem[{\citenamefont{Kostelecky and Mewes}(2002)}]{Kostelecky:2002}
\bibinfo{author}{\bibfnamefont{V.~A.} \bibnamefont{Kostelecky}}
  \bibnamefont{and} \bibinfo{author}{\bibfnamefont{M.}~\bibnamefont{Mewes}},
  \bibinfo{journal}{Phys. Rev. D} \textbf{\bibinfo{volume}{66}},
  \bibinfo{pages}{056005} (\bibinfo{year}{2002}).

\bibitem[{\citenamefont{Saathoff et~al.}(2003)\citenamefont{Saathoff, Karpuk,
  Eisenbarth, Huber, Krohn, Horta, Reinhardt, Schwalm, Wolf, and
  Gwinner}}]{Saathoff:2003}
\bibinfo{author}{\bibfnamefont{G.}~\bibnamefont{Saathoff}},
  \bibinfo{author}{\bibfnamefont{S.}~\bibnamefont{Karpuk}},
  \bibinfo{author}{\bibfnamefont{U.}~\bibnamefont{Eisenbarth}},
  \bibinfo{author}{\bibfnamefont{G.}~\bibnamefont{Huber}},
  \bibinfo{author}{\bibfnamefont{S.}~\bibnamefont{Krohn}},
  \bibinfo{author}{\bibfnamefont{R.~M.} \bibnamefont{Horta}},
  \bibinfo{author}{\bibfnamefont{S.}~\bibnamefont{Reinhardt}},
  \bibinfo{author}{\bibfnamefont{D.}~\bibnamefont{Schwalm}},
  \bibinfo{author}{\bibfnamefont{A.}~\bibnamefont{Wolf}}, \bibnamefont{and}
  \bibinfo{author}{\bibfnamefont{G.}~\bibnamefont{Gwinner}},
  \bibinfo{journal}{Phys. Rev. Lett.} \textbf{\bibinfo{volume}{91}},
  \bibinfo{pages}{190403} (\bibinfo{year}{2003}).

\bibitem[{\citenamefont{Robertson}(1949)}]{Robertson:1949}
\bibinfo{author}{\bibfnamefont{H.~P.} \bibnamefont{Robertson}},
  \bibinfo{journal}{Rev. Mod. Physics} \textbf{\bibinfo{volume}{21}},
  \bibinfo{pages}{378} (\bibinfo{year}{1949}).

\bibitem[{\citenamefont{Mansouri and Sexl}(1977)}]{Mansouri:1977}
\bibinfo{author}{\bibfnamefont{R.}~\bibnamefont{Mansouri}} \bibnamefont{and}
  \bibinfo{author}{\bibfnamefont{R.~U.} \bibnamefont{Sexl}},
  \bibinfo{journal}{Gen. Rel. Grav.} \textbf{\bibinfo{volume}{8}},
  \bibinfo{pages}{497} (\bibinfo{year}{1977}).

\end{thebibliography}

\end{document}